# Women in physics in the United States: recruitment and retention

Nina Abramzon[1], Patrice Benson[2], Edmund Bertschinger[3], Susan Blessing[4], Geraldine L. Cochran[5], Anne Cox[6], Beth A. Cunningham[7], Jessica Galbraith-Frew[8], Jolene Johnson[9], Leslie Kerby[10], Elaine Lalanne[11], Christine O'Donnell[12], Sara Petty[13], Sujatha Sampath[14], Chandralekha Singh[15], Cherrill Spencer[16], Kathryne Sparks Woodle[17], and Sherry Yenello[18]

[1]California Polytechnic University Pomona; [2]US Military Academy; [3]Massachusetts Institute of Technology; [4]Florida State University; [5]Rochester Institute of Technology; [6]Eckerd College; [7]American Association of Physics Teachers; [8]University of Utah; [9]St. Catherine University; [10]University of Idaho; [11]US Navy; [12]University of Virginia; [13]Virginia Polytechnic Institute and State University; [14]University of Wisconsin-Milwaukee; [15]University of Pittsburgh; [16]SLAC National Accelerator Laboratory; [17]Penn State University; [18]Texas A&M University

a) Corresponding Author: glcsps@rit.edu

**Abstract.** Initiatives to increase the number, persistence, and success of women in physics in the US reach pre-teen girls through senior women. Programs exist at both the local and national levels. In addition, researchers have investigated issues related to gender equity in physics and physics education. Anecdotal evidence suggests increased media coverage of the underrepresentation of women in science. All of these efforts are both motivated and made more effective by the collection and presentation of data on the presence, persistence, and promise of women in physics.

Although the numbers of women in physics in the US at all levels have been increasing, the fraction of women involved in physics is still low. Moreover, the more senior levels of physics positions are still held primarily by men (8% full professor [1]). The underrepresentation of women in physics, as well as broader issues for women in all science, technology, engineering, and mathematics fields (STEM), is receiving wider societal attention through recent articles in major newspapers[2] and news media in the US[3]. Efforts to address these issues through improvement in the recruitment and retention of women in physics in the US span a wide range of activities, from encouraging girls to consider a future in STEM to professional development for academic women physicists.

Early interventions include nationally held conferences and programs aimed at young girls. As an example, Expanding Your Horizons (EYH) in Science, Engineering, and Mathematics conferences are one-day programs for girls aged 12 to 18. EYH conferences engage young girls in interesting and fun hands-on STEM activities. These activities, each lasting one hour, are designed and led by professional women in STEM fields. Anecdotal evidence and several surveys show that attending an EYH sparks girls' interest and overcomes their prejudices about STEM subjects[4]. The EYH conference format was started in 1976 in California. Since then about 900,000 girls have attended an EYH, which are now held annually in 85 cities in 37 US states and several other countries. The EYH Network[5] helps new conference sites to get started and provides technical assistance.

Aiming at young adults, the American Physical Society (APS) hosts Conferences for Undergraduate Women in Physics (CUWiP)[6]. These events are three-day regional conferences primarily for undergraduate physics majors. Beginning in 2006 with one site and 29 attendees, the annual conferences now reach across 8 sites hosting more than 1,000 students each year. The goal of the CUWiPs is to help undergraduate women continue in physics by providing them with information about graduate school and professions in physics, the opportunity to experience a professional conference, and a network of other women in physics. A typical program includes research talks, panel discussions about a broad range of careers in physics and graduate school, presentations and discussions about women in physics, lab tours, and a student poster session. In their evaluation of the 2014 CUWiP, Brewe and Hazari[7] found that the conferences resulted in consistent and strongly positive understanding of physics careers and how to pursue them. Further, the conferences led to a growth in community and development of mentors among conference participants.

Workshops to aid in the retention and promotion of women faculty in physics are held at the meetings of the APS and the American Association of Physics Teachers[8]. The workshops are designed to provide women with professional development and support in navigating careers in physics. Workshop topics include gender differences in views on competition, strategies for engaging in successful negotiations, and tactics to improve communication and resolve difficult situations and conflicts at the workplace for women physicists. These workshops are interactive and include the results of recent research relevant to the topics discussed.

On March 11, 2009, President Obama signed an Executive Order creating the White House Council on Women and Girls[9] with the purpose of ensuring that each federal agency takes into account the needs of women and girls in their policies, programs, and legislation. The office of Science and Technology Policy has a "Women in STEM" program[10] that provides fact sheets with information and statistics on women and girls in STEM and resources including college and career readiness information, Title IX compliance (prohibition of gender discrimination), mentoring, and workplace flexibility. The National Science Foundation, a US government scientific funding agency has funded ADVANCE[11], an interdisciplinary program to advance careers of women in science and engineering, specifically in higher education. An important aspect of the ADVANCE program is the effort to shift from "fixing the women" to transforming institutional structures to reduce the barriers to the advancement of women faculty in science and engineering fields[12].

In recent years physics education research (PER) has included gender studies that investigate why female students choose to study physics or pursue careers in physical science and how to help female students be successful in their physics studies. For example, using a logistic regression model to predict the passing of students in an introductory physics with calculus course, Sawtelle et al.[13] found that the probability of passing the course relies on self-efficacy opportunities that are different for men and women. Miyake et al.[14] investigated the use of values affirmation, a psychological intervention, in college-level introductory physics classes and found that the intervention reduced the male-female performance and learning difference substantially. Kost et al.[15] found that gender gaps on post-test scores in introductory physics courses were associated with factors determined prior to taking the course, such as previous physics and math knowledge and incoming attitudes and beliefs. Using multivariate matching methods, Hazari et al.[16] determined factors that had significant effects on female students' decisions to pursue careers in physical science. Results of these, and other, studies in PER have implications on pedagogical strategies and classroom dynamics in physics classroom at the pre-college and collegiate levels.

This work was supported by the National Science Foundation under grant #PHY-1419453.